\documentclass[aip,jap,preprint,showpacs,eqsecnum,floatfix]{revtex4-1}

\usepackage{graphicx} 

\usepackage[english]{babel}

\usepackage{textcomp}

\usepackage{microtype} 

\begin{document}

\title{Raman spectroscopy on etched graphene nanoribbons}
\author{D. Bischoff$^{1}$, J. G\"uttinger$^1$, S. Dr\"oscher$^1$, T. Ihn$^1$, K. Ensslin$^1$ and C. Stampfer$^{1,2}$}
\affiliation{	$^1$Solid State Physics Laboratory, ETH Zurich, 8093 Zurich, Switzerland \\
		$^2$JARA-FIT and II. Institute of Physics, RWTH Aachen, 52074 Aachen, Germany}
\date{\today}


\begin{abstract}
We investigate etched single-layer graphene nanoribbons with different widths ranging from 30 to 130 nm by confocal Raman spectroscopy. We show that the D-line intensity only depends on the edge-region  of the nanoribbon and that consequently the fabrication process does not introduce bulk defects. In contrast, the G- and the 2D-lines scale linearly with the irradiated area and therefore with the width of the ribbons. We further give indications that the D- to G-line ratio can be used to gain information about the crystallographic orientation of the underlying graphene. Finally, we perform polarization angle dependent measurements to analyze the nanoribbon edge-regions. 
\end{abstract}

\maketitle
\newpage

\section*{Introduction}

Graphene nanoribbons~\cite{han07,lin08,wan08,mol09b,han09,mol10,gal10} attract increasing attention due to the possibility of building graphene-based nanoelectronics as for example field-effect transistors~\cite{wan08,zha08} or quantum dot devices~\cite{sta08a,pon08}. In contrast to two-dimensional gapless bulk graphene~\cite{gei07}, it has been shown that confinement~\cite{fer07a,yan07a}, disorder~\cite{sol07} and edge effects~\cite{yan07a} introduce a transport gap in graphene nanoribbons.
The fabrication technique may influence the transport properties of the nanoribbons in terms of added bulk and/or edge disorder. Disorder is expected to strongly influence the scaling behavior of the energy gap as a function of the nanoribbon width and the local doping profile~\cite{mol09b,han09,mol10,gal10}. In addition, very little is known about the edge structure and theoretical investigations of the vibrational properties of nanoribbons have only been started very recently~\cite{gil09}.
Raman spectroscopy on carbon (nano)materials~\cite{mal09,fer07} has been recognized as a powerful technique not only for probing selected phonons, but also for identifying the number of graphene layers~\cite{fer06,dav07a}, for determining local doping levels~\cite{sta07aa}, for studying electron-phonon coupling~\cite{pis07} and thus for the electronic properties themselves.

In this letter we report Raman spectroscopy experiments on etched graphene nanoribbons with different widths $w$ ranging from 30 to 130~nm (schematic in Fig.~1a). We show that the characteristic signatures of single-layer graphene (SLG) in the Raman spectra are still well preserved, that the absolute G- and 2D-line intensities scale with the nanoribbon width, whereas the D-line intensity does not. Consequently, the D-line intensity depends only on the edge-region of the nanoribbon including the edge  roughness which can be further analyzed by performing polarization dependent measurements. 

\section*{Fabrication}

The nanoribbon fabrication is based on the mechanical exfoliation of natural graphite~\cite{nov04}, electron beam lithography, reactive ion etching and metal evaporation. For details see Ref.~\cite{mol09b}. The graphene flakes have been identified to consist of a  single-layer by measuring the Raman full width at half maximum (FWHM) of the 2D-line prior to processing~\cite{fer06,dav07a}. In Fig.~1b we show a scanning force microscope (SFM) image of six etched graphene nanoribbons with different widths. In total, we have studied three such nanoribbon arrays fabricated from three different single-layer flakes resulting in a total of more than 20 individual nanoribbons.

\section*{Experimental Setup}
All Raman spectra were acquired using a green laser (532~nm, $\hbar \omega_L=$ 2.33 eV). Employing a long working distance focusing lens (numerical aperture~=~0.80), we obtain a spot size with a diameter $d~\approx$ 400~nm. The incident light was -- unless stated differently -- linearly polarized parallel to the macroscopic edge of the nanoribbons ($\theta=0^\circ$) and detection was always insensitive to polarization. The laser power was set to $\approx$ 2~mW in order to exclude heating effects and all measurements were conducted at room temperature.

\begin{figure}[bt]
	\centering %
	\includegraphics[width=0.5\linewidth]{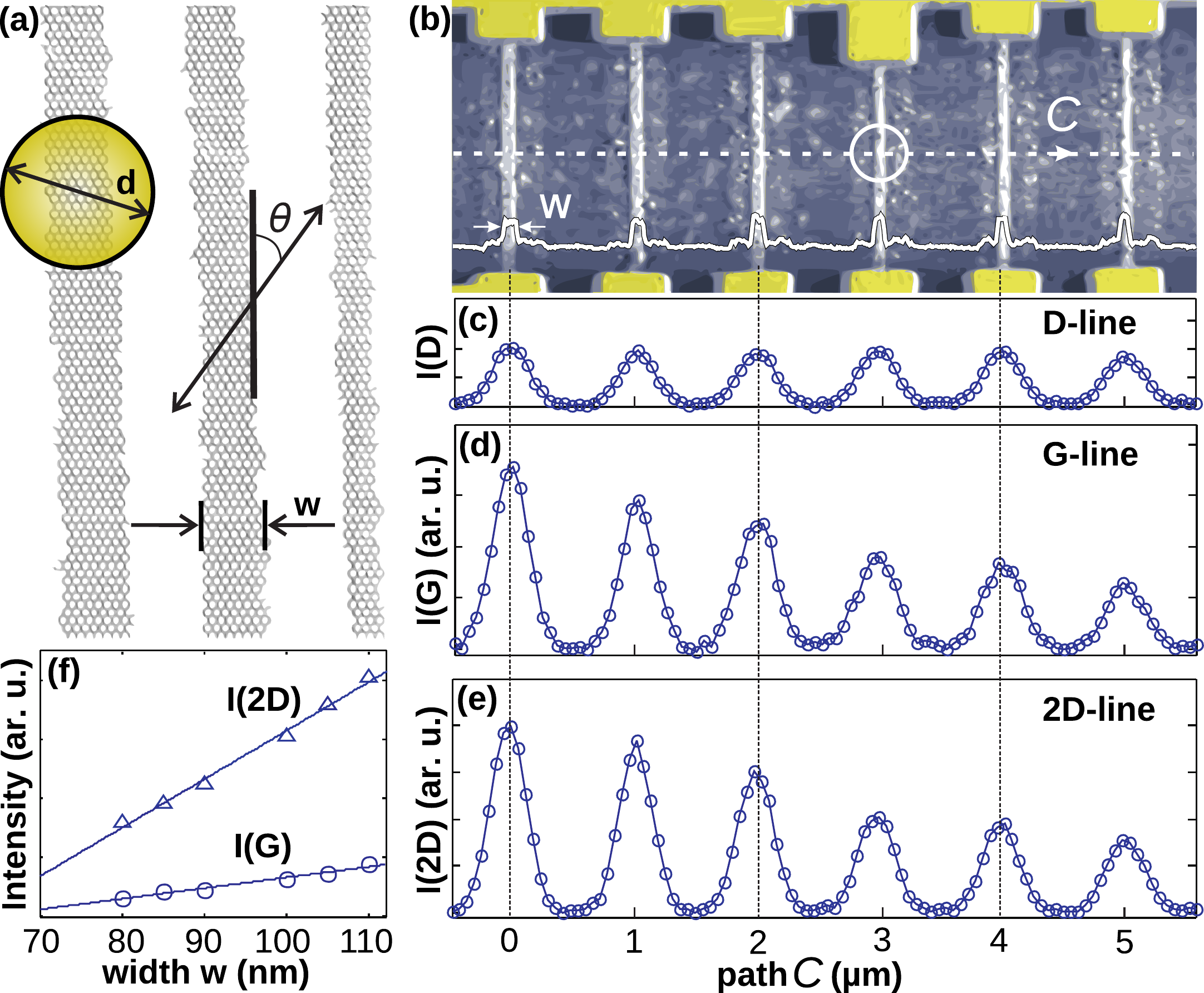}%
	\caption[FIG1]{%
(color online) %
(a) Schematic illustration of etched nanoribbons with widths $w$. The yellow circle represents the Raman laser spot of the linearly polarized light with angle $\theta$.
(b) SFM image of six etched graphene nanoribbons (vertical white bars) with corresponding gold contacts (yellow) which were used to align and identify the nanoribbons during the Raman measurements. The nanoribbon width $w$ decreases from left to right from 110~nm down to 80~nm and was determined by the SFM cross-section shown in the lower part of the image (white line). The laser spot-size is represented by the white circle. 
(c) Raman line scan along path $C$ [see panel (b)]. For each data point, a Raman spectrum was recorded and the intensity of the D-line was calculated by summing up the detector counts (from 1306-1375 cm$^{-1}$).
(d) Same as above, but for the G-line (1558-1609 cm$^{-1}$) and 
(e) the 2D-line (2618-2734 cm$^{-1}$).
(f) Maximum intensity of the G- and the 2D-line plotted versus the nanoribbon width.%
} 
	\label{fig:LineScan}
\end{figure}

\begin{figure}[t]
	\centering %
	\includegraphics[width=0.5\linewidth]{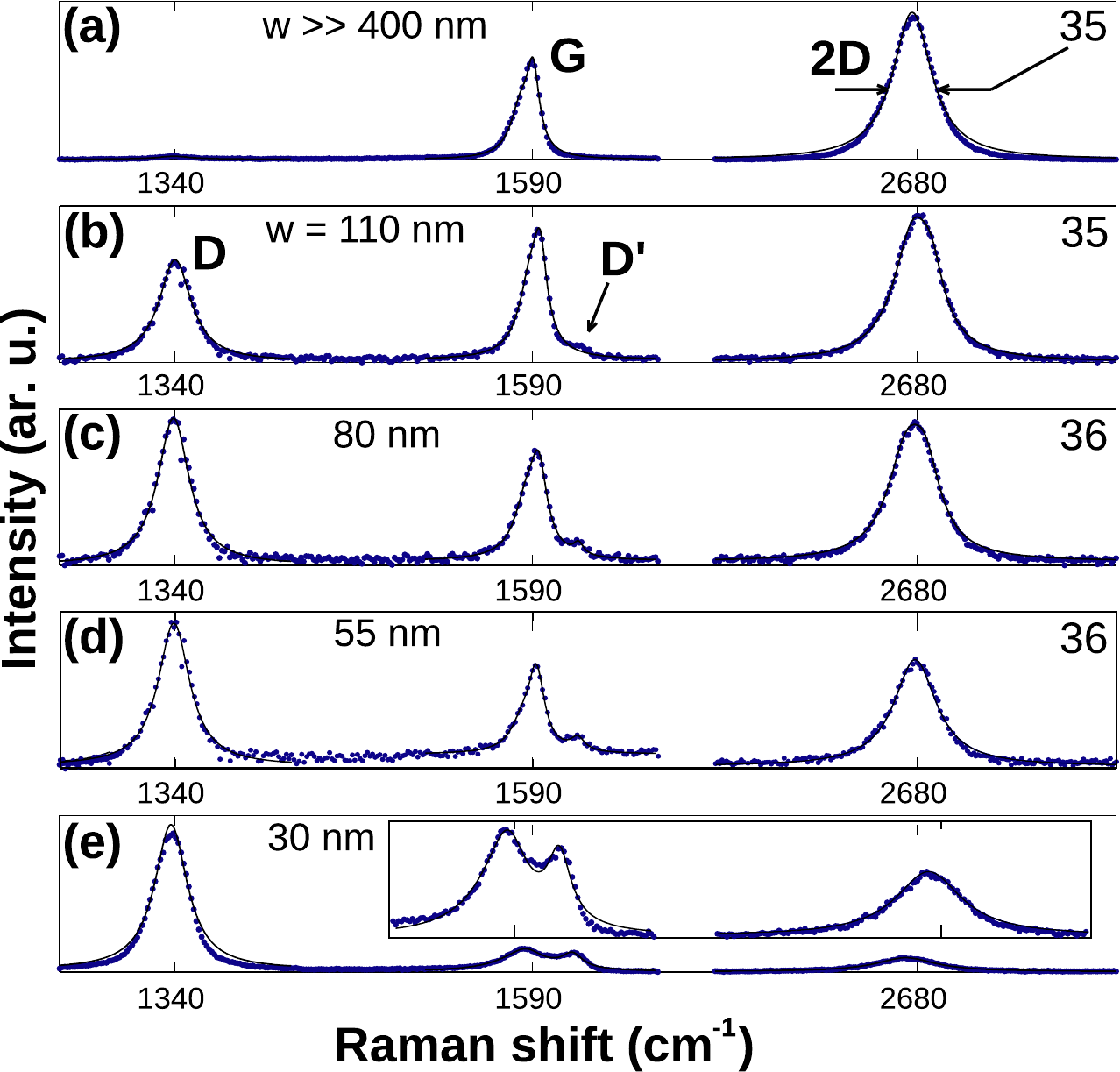}
	\caption[FIG2]{%
(color online) %
A selection of Raman spectra from nanoribbons of different widths. For each spectrum, the silicon background signal was removed. 
In the inset of panel (e), magnified versions of G- and 2D-lines are shown. The blue dots represent measured data points whereas the black lines represent best fits.
	} 
	\label{fig:SpectraOverview}
\end{figure}

\section*{Results}

A selection of Raman spectra of nanoribbons with different widths $w$ is shown in Fig.~2. In all measurements, the characteristic signatures of graphene Raman spectra can be well identified: The defect induced D-line ($\approx$1340~cm$^{-1}$), the G-line ($\approx$1580-1590~cm$^{-1}$), the 2D-line ($\approx$2680~cm$^{-1}$) and for the thinner ribbons an additional defect induced D'-line ($\approx$1620~cm$^{-1}$). For a detailed review on the microscopic nature of all these lines, see Refs.~\cite{tho00,rei04,fer06,dav07a,fer07,pis07,mal09}.

The FWHM of the 2D-line with values below 40~cm$^{-1}$ (see right labels in Figs.~2a-d) is a strong signature for the single-layer nature of the graphene nanoribbons~\cite{fer06,dav07a}. In order to enhance the signal for thin nanoribbons, we recorded the spectra of the 30~nm nanoribbon on a sub-array of several closely spaced nanoribbons with equal width (approx. 6 ribbons irradiated by the laser). This sample averaging may also explain the rather large but still single-layer FWHM of the 2D-line of nearly 50~cm$^{-1}$. In contrast to defectless bulk graphene where significant D- and D'-lines are rarely present, we observe strong D- and D'-lines in the recorded spectra as the edges act as defects and allow elastic inter-valley scattering of electrons~\cite{can04,can04a,cas09}.

As demonstrated in Fig.~1c, the D-line intensity does not depend on the width of the nanoribbons. The peak width in Figs.~1c-e arises from strong spatial oversampling as the laser spot size ($\approx$ 400~nm) is significantly larger than the step size ($\approx$ 70~nm). As the laser spot size is also significantly larger than the nanoribbon width ($d\gg w$), the total irradiated edge length is approximately $2\times d$ (left and right edge) -- independent of the nanoribbon width. As long as there is no bulk disorder present, the intensity of the D-line is not expected to depend on the nanoribbon width. This is observed in Fig.~1c and leads to the important conclusion that the reactive ion etching process used to pattern nanoribbons does not introduce a detectable amount of bulk defects into our graphene nanostructures.

In contrast to the D-line, we observe a width dependence of the G- and 2D-line intensities as shown in Figs.~1d,e. The intensity of the G-line is a function of the amount of irradiated sp$^2$ bound carbon atoms~\cite{mal09} and therefore expected to directly depend on the graphene nanoribbon area below the laser spot ($\approx d\times w$). The same proportionality also holds for the 2D-line~\cite{tui70,cas01,Laz08}. This expected linear dependence of the G- and 2D-line heights is confirmed in Fig.~1f (see lines fitted to circles and triangles). It is crucial to note that Fig.~1f does not allow to determine the width of a possibly existing Raman-inactive edge-region~\cite{han07}. This is due to offsets in Figs.~1c-e (background of the Si substrate) and due to the inherent noise in the measurement, which masks the small line heights resulting from the short integration time. From the fact that we can measure 30~nm wide nanoribbons (see Fig. 2e), we conclude that such a Raman-inactive edge-region  must either be significantly smaller than 15~nm or scale with the nanoribbon width.

\begin{figure}[t]
	\centering%
	\includegraphics[width=0.5\linewidth]{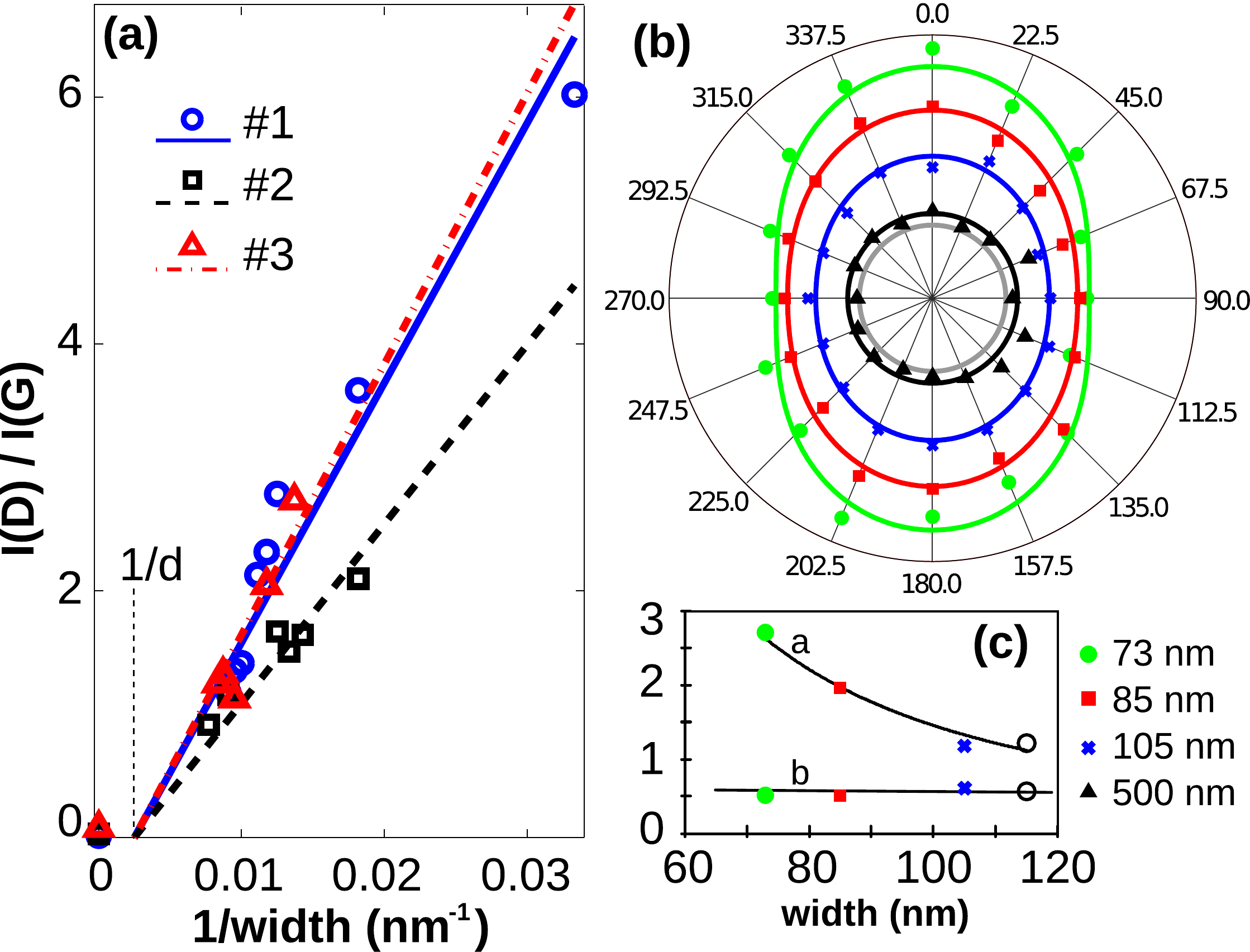}
	\caption[FIG3]{ %
(color online) %
(a) %
Intensity ratio $I(D)/I(G)$ versus $w^{-1}$ for different nanoribbons located on the three different arrays ($\#1-\#3$). 
(b) %
Polarization angle dependence of I(D)/I(G) for nanoribbons of different width. The inner gray circle is located at I(D)/I(G)=0 and the outer at I(D)/I(G) = 3.25. The data points are fitted with $I(D)/I(G)(\theta)=a[b+(1-b)\cos^2\theta]$.
(c) %
Ribbon width dependence of the fit parameters $a$ and $b$.
	}
	\label{fig:Orientation}
\end{figure}


In order to compare the three arrays ($\#1-\#3$), we show in Fig.~3a the intensity ratio $I(D)/I(G)$ versus the inverse ribbon width. Each array is made from a single SLG flake and all nanoribbons on a specific array were etched in the same processing step and are oriented parallel to each other. As a measure for the intensity, the peak area was used~\cite{com0f}. Based on the previous argument, the relation $I(D)/I(G) \propto w^{-1}$  is expected and observed. It is important to note that this relation only holds in the limit $d\gg w$ (otherwise there is no edge located under the laser spot). Therefore, a linear fit was performed for each array and the intersection of the fit with the abscissa  was pinned to $w^{-1}=1/d$.

Interestingly, the fits for the different arrays exhibit different slopes. There are two possible explanations: (i) The edge roughness differs for the different arrays: Rougher edges result in more defects illuminated by the laser spot and therefore a higher D-line intensity. As all the arrays were fabricated using the same process, this is improbable. (ii) The slopes can also be attributed to different crystallographic orientations of the graphene relative to the edge. This is because perfect zig-zag edges do not activate the D-line due to momentum conservation~\cite{can04,can04a}, whereas armchair segments do. While the plasma etched edges are certainly rough, the ratio of zig-zag to armchair segments should nevertheless depend on the overall crystallographic orientation relative to the nanoribbon orientation.


In Fig.~3b, we show the $I(D)/I(G)$ ratio for different nanoribbons of one array as a function of the incident photon polarization angle  $\theta$. For each ribbon and each angle, a spectrum was recorded and both the G- and the D-line were fitted each with a single Lorentzian. Geometric considerations suggesting mirror planes at 0$^{\circ}$ and 90$^{\circ}$ are confirmed. For Raman measurements in graphene, only phonon wave vectors perpendicular to the incident light polarization contribute to the signal~\cite{gru03,can04a}. As phonon wave vectors along the principal nanoribbon axis (i.e. $\theta = $90$^{\circ}$ and 270$^{\circ}$) see little of the edges, the D-line is expected to be more suppressed than at all other angles.

Following Casiraghi et al.~\cite{cas09} describing similar measurements on graphene edges and taking into account that $I(G)$ does not depend on $\theta$~\cite{com0c}, the following expression can be derived: $I(D)/I(G)(\theta)=a[b+(1-b)\cos^2\theta]$, where $a=I(D)_{max}/I(G) \propto 1/w$ and $b=I(D)_{min}/I(D)_{max}$. This relation was used to fit the data and indeed a strong width dependence of $a\propto 1/w$ and an overall constant of $b \approx$ 0.55 only depending on the structure of the edges are found (Fig.~3.c). According to Ref.~\cite{cas09}, a lower bound of the edge disorder correlation length can be estimated by $\xi=2 v_F/(\omega_L b)\approx$ 1~nm, which is in reasonable agreement with current fabrication limitations~\cite{com0a}.

\section*{Conclusion}
In conclusion, we have demonstrated that the D- and D'-line depend only on the nanoribbon edge-region  whereas the G- and the 2D-line scale with the illuminated area. We have shown that our fabrication process does not introduce bulk defects and that the $I(D)/I(G)$ ratio can give indications about the crystallographic orientation of graphene. These insights may help in designing further experiments and designing future graphene nanoelectronic devices.


\section*{Acknowledgment}
The authors thank C. Casiraghi, A. C. Ferrari, M. Haluska, A. Jorio and L. Wirtz for helpful discussions and C. Hierold for providing access to the Raman spectrometer. Support by SNSF and NCCR nanoscience is gratefully acknowledged.



\begin{thebibliography}{99}
\bibitem{com0f}
The area can be more accurately determined than the height and the area represents all photons/phonons participating in the Raman process.
\bibitem{com0c}
It is hard to prove that the G-line intensity is independent of $\theta$, as even the slightest change in the focus of the Raman spectrometer changes the signal intensity. We therefore always use ratios when comparing line intensities.
\bibitem{com0a}
Here we assume that the length of the local armchair segments is significantly larger than the carbon-carbon bond length. The disordered edge direction somewhere between averaged armchair or zigzag direction is not known and therefore only a very rough estimate on the lower bound can be made \cite{cas09}.
\end{thebibliography}
\end{document}